\documentclass[showpacs,preprintnumbers,amsmath,amssymb,prl,twocolumn,floatfix,amssymb]{revtex4}
\usepackage[dvips]{graphicx,color}
\usepackage{dcolumn}
\usepackage{amsmath,amssymb,bbold,bm}

\begin{document}
\title{Phonon-induced topological transitions and crossovers in Dirac materials}
\author{Ion Garate}
\affiliation{Department of Physics, Yale University, New Haven, CT 06520, USA}
\date{\today}
\begin{abstract}
We show that electron-phonon interactions can alter the topological properties of Dirac insulators and semimetals, both at zero and nonzero temperature.
Contrary to the common belief that increasing temperature always destabilizes topological phases, our results highlight instances in which phonons may lead to the appearance of topological surface states above a crossover temperature in a material that has a  topologically trivial ground state.
\end{abstract}
\maketitle

{\em Introduction.---}
Recent years have witnessed the unexpected discovery of novel topological phases in certain electrically insulating~\cite{ti} and semimetallic~\cite{ws} solids whose low-energy bulk excitations are Dirac fermions.
Topological phases in these ``Dirac materials'' are characterized by nonzero integers (topological invariants), which manifest themselves through peculiar and robustly gapless states localized at the boundaries of the material. 
Partly enticed by a vision of transistors that would operate by switching topological invariants on and off, there is keen interest in finding ways to induce topological phases in intrinsically non-topological materials.

In centrosymmetric crystals, quantum phase transitions between topological and non-topological insulators occur when 
the sign of the Dirac fermions' mass gets inverted at an odd number of time-reversal-invariant momenta (TRIM) in the bulk Brillouin zone.
In equilibrium, such ``band inversions'' can result from applying pressure~\cite{bahramy2012} or dc electric fields~\cite{kim2012}, from changing the compound stoichiometry~\cite{xu2011}, or from strong alloying~\cite{guo2010}.
Away from equilibrium, ac electromagnetic fields may effectively produce a band inversion~\cite{lindner2011},
and engineered coupling to dissipative baths presumably induces topological phases in driven cold atomic systems~\cite{diehl2011}.
The latter two cases illustrate how the topological invariant for an electron system may be altered after coupling it to a non-electronic environment, and motivate the question of whether analogous environment-induced effects could occur {\em in equilibrium}.

In crystals that are in thermodynamic equilibrium, lattice vibrations form the most ubiquitous bath that couples to electrons.
Although electron-phonon coupling can hardly be engineered in a given material, its influence on electronic properties is strongly temperature-dependent.
Even as a recent formal study suggests that topological invariants are not robust under dephasing~\cite{avron2011}, 
a growing body of work~\cite{huang2008,hatch2011,cheng2011,zhu2011,giraud2011,thalmeier2011,budich2012,pan2012} treats phonons as mere actors that do not alter the topology of the underlying electronic stage.
The objective of our paper is to demonstrate that (i) phonons can modify topological properties of an electronic structure, and (ii) temperature can be used as the ``agent'' that drives those changes. 

Electron-phonon interactions are known to often decrease the bandgap of ordinary semiconductors monotonically as a function of temperature~\cite{allen1981}.
However, it has not been recognized that, at least in narrow-gap semiconductors, phonon-induced renormalization of the bandgap can culminate in a band inversion both at zero and nonzero temperature. 
In Dirac insulators with non-topological ground states, a band inversion arising at nonzero temperature implies the emergence of peculiar gapless surface states {\em above} that temperature.
This may be regarded as a ``topological thermal crossover'', which has not been previously discussed.

{\em Electron-phonon self-energy.--}
Consider electrons in a perfect periodic lattice with crystal Hamiltonian $h^{(0)}({\bf k})$, whose energy eigenvalues $E_{{\bf k}\alpha}$ and eigenstates $|\Psi_{{\bf k}\alpha}\rangle\equiv\exp(i{\bf k}\cdot{\bf r})|{\bf k}\alpha\rangle/V^{1/2}$ characterize the electronic structure. 
Here $V$ is the volume of the crystal~\cite{expansion}, ${\bf r}$ is the position vector, ${\bf k}$ is the crystal momentum, $\alpha=1,...,N$ is the band index and $|{\bf k}\alpha\rangle$ is a $N$-component eigenspinor.
The change in the electronic structure due to lattice vibrations is approximately captured by the electron-phonon self-energy~\cite{mahan,dw}
\begin{align}
\label{eq:se}
&\Sigma_{\alpha\alpha'}(i\omega_n,{\bf k})=\sum_{{\bf q},\beta} g_{\bf q}^2 \langle{\bf k}\alpha|{\bf k}-{\bf q}\beta\rangle\langle{\bf k}-{\bf q}\beta|{\bf k}\alpha'\rangle \nonumber\\
~~&\times\left[\frac{1+n_{\bf q}-f_{{\bf k-q}\beta}}{i\omega_n-\xi_{{\bf k-q}\beta}-\omega_{\bf q}}+\frac{n_{\bf q}+f_{{\bf k-q}\beta}}{i\omega_n-\xi_{{\bf k-q}\beta}+\omega_{\bf q}}\right],
\end{align}
where $\omega_n\equiv(2 n+1)\pi T$ ($n\in\mathbb{Z}$) is the Matsubara frequency, $T$ is the temperature, $\xi_{{\bf k}\alpha}=E_{{\bf k}\alpha}-\epsilon_F$, $\epsilon_F$ is the Fermi energy, $\omega_{\bf q}$ is the phonon dispersion, $n_{\bf q}=[\exp(\omega_{\bf q}/T)-1]^{-1}$ is the phonon occupation number, $f_{{\bf k}\alpha}=[\exp(\xi_{{\bf k}\alpha}/T)+1]^{-1}$ is the fermion occupation number, and $g_{\bf q}$ is the electron-phonon coupling.
For concreteness and simplicity we concentrate on the deformation coupling to longitudinal acoustic phonons~\cite{mahan}, so that  $g_{\bf q}^2=\hbar C^2 q/(2\rho V c_s)$ and $\omega_{\bf q}=c_s q$, where
$C$ is the deformation potential, $\rho$ is the atomic mass density and $c_s$ is the sound velocity.

In presence of lattice vibrations, the electronic Green's function $G$ obeys $G_{\alpha\alpha'}^{-1}(i\omega_n,{\bf k})=\left(i\omega_n-\xi_{{\bf k}\alpha}\right)\delta_{\alpha\alpha'}-\Sigma_{\alpha\alpha'}(i\omega_n,{\bf k})$, where $\delta$ is Kronecker's delta.
The off-diagonal matrix elements of the self-energy are generally nonzero and lead to avoided band crossings except at certain (e.g. topologically protected) degeneracy points.
Upon diagonalization, the Green's function obeys  $G_{\alpha\alpha'}^{-1}(i\omega_n,{\bf k})=\left(i\omega_n+\epsilon_F-\chi_\alpha(i\omega_n,{\bf k})\right)\delta_{\alpha\alpha'}$,  and the solutions of
\begin{equation}
\label{eq:heff}
E^*_{{\bf k}\alpha}+i\Gamma_{{\bf k}\alpha}=\chi_\alpha(E^*_{{\bf k}\alpha}-\epsilon_F+i\Gamma_{{\bf k}\alpha},{\bf k})
\end{equation}
yield the energy dispersion $E^*_{{\bf k}\alpha}$ and broadening $|\Gamma_{{\bf k}\alpha}|$ of dressed quasiparticles at temperature $T$.

{\em Phonon-induced topological transitions in insulators.--}
A minimal lattice model that captures the generic low-energy physics of 3D Dirac insulators reads~\cite{gili2010} 
\begin{equation}
\label{eq:model}
h^{(0)}({\bf k})= \epsilon_{0,{\bf k}}{\bf 1}_4+ {\bf d}_{\bf k}\cdot{\boldsymbol\sigma}\tau^x+M_{\bf k}{\bf 1}_2\tau^z,
\end{equation}
where ${\bf 1}_n$ is an $n\times n$ identity matrix, $\sigma^i$ and $\tau^i$ are Pauli matrices in ``spin'' and ``orbital'' space (respectively), $\epsilon_{0,{\bf k}}=2\gamma\left(3-\sum_i\cos(k_i a)\right)$, $d_{i,{\bf k}}=-2\lambda\sin(k_i a)$, $M_{\bf k}=m+2 t\left(3-\sum_i \cos(k_i a)\right)$, $i\in\{x,y,z\}$, $a$ is the lattice constant,  $(\gamma,\lambda,t)$ are band parameters and $m$ is the Dirac mass.
This model describes an insulator with doubly degenerate bands and an energy gap $E_g= 2 m$ at $k=0$.
Anticipating that $|m|\lesssim |t|$ in typical Dirac materials, the insulator is topological if $m\, t<0$ and non-topological if $m\, t>0$.
Since $|E_{{\bf k}\alpha}|\gg|m|$ for $k\neq 0$ TRIM, phonon-induced changes in topological properties are realized via band inversions at $k=0$. 
Thus we focus on $\Sigma_{\alpha\alpha'}(i\omega_n,{\bf 0})$, which can be recasted in matrix form as
\begin{equation}
\label{eq:se2}
\hat{\Sigma}(i\omega_n,{\bf 0})=\Sigma_0(i\omega_n){\bf 1}_4+\Sigma_z(i\omega_n){\bf 1}_2\tau^z,
\end{equation}
with $\Sigma_0(i\omega_n)\equiv(\Sigma_{1 1}(i\omega_n,{\bf 0})+\Sigma_{3 3}(i\omega_n,{\bf 0}))/2$ and $\Sigma_z(i\omega_n)\equiv(\Sigma_{1 1}(i\omega_n,{\bf 0})-\Sigma_{3 3}(i\omega_n,{\bf 0}))/2$. 

At $T=0$, a topological quantum phase transition occurs when the renormalized Dirac mass, $m^*\equiv m+\Sigma_z(0)$, crosses zero and changes sign~\cite{wang2012}.
Due to the frequency-dependence of the self-energy, $2 m^*$ is in general {\em not equal} to the renormalized energy gap $E_g^*\equiv E^*_c-E^*_v$,
where $E^*_c\simeq m+\Sigma_0(E^*_c-\epsilon_F)+\Sigma_z(E^*_c-\epsilon_F)$ and $E^*_v\simeq -m+\Sigma_0(E^*_v-\epsilon_F)-\Sigma_z(E^*_v-\epsilon_F)$.
However, for the toy model and range of parameters taken here, $\Sigma_{\alpha\alpha'}(\omega,{\bf k})\simeq\Sigma_{\alpha\alpha'}(0,{\bf k})$ ~\cite{freq} and thus 
the sign change of $m^*$ occurs approximately simultaneously with the band inversion of the dressed bulk quasiparticle spectrum.

One interesting aspect of phonons is that they make $m^*$ strongly $T$-dependent on the scale of the Debye temperature.
In contrast, the Dirac-mass renormalization due to static disorder~\cite{guo2010} is $T$-independent, and the renormalization due to purely electronic Coulomb interactions is only weakly $T$-dependent~\cite{ee}.
Provided that quasiparticle broadening is small compared to $|m^*(T)|$, the topological properties of $G$ are identical to those of $h^{(0)}$ with $m\to m^*(T)$.   
Therefore, a sign change of $m^*(T)$ at $T=T_c$ implies the emergence (if $m^*(0) t >0$) or evanescence (if $m^*(0) t <0$) of topological surface states at $T\gtrsim T_c$.
Next, we show that phonons may induce such a ``thermal topological crossover''. 

 
Adopting the spherical approximation of Eq.~(\ref{eq:model}) near ${\bf k}={\bf 0}$ and introducing a UV momentum cutoff $k_c=\pi/a$, Eq.~(\ref{eq:se2}) can be evaluated analytically in the limits  $T\ll\omega_c$ and $T\gg\omega_c$, where $\omega_c\equiv c_s k_c$ is the Debye frequency.
We obtain
\begin{align}
\label{eq:se3}
\Sigma_{0,z}=& U^2 f_{0,z}(T), 
\end{align}
where $U\equiv (C/a^2)(\hbar/2\rho c_s)^{1/2}$ is a constant in units of energy, and $f_{0,z}(T)$ are functions of electronic structure parameters and of temperature, with 
\begin{align}
\label{eq:fz1}
f_z (T\ll\omega_c) \simeq & -y\left[1-x^2\ln(1+1/x^2)\right]/4+O(T^4)\nonumber\\
f_z (T\gg\omega_c) \simeq & -(T/\omega_c) y\left[1-x\tan^{-1}(1/x)\right],
\end{align}
$x\equiv (2/\pi)\lambda/(t^2-\gamma^2)^{1/2}$ and $y\equiv t/(t^2-\gamma^2)$. 
Similarly, $f_0(T)=-(\gamma/t) f_z(T)$. 
To leading order, Eq.~(\ref{eq:fz1}) is independent of $m$; the subleading $m$-dependent terms become negligible when $|m|\ll|t|$.
In the derivation of Eq.~(\ref{eq:fz1}) we have neglected $\omega_{\bf q}$ from the denominators of Eq.~(\ref{eq:se}), which is a good approximation as confirmed numerically. 
This approximation leads to the cancellation of Fermi factors in Eq.~(\ref{eq:se}) and thus highlights the fact that $\omega_c$ governs the $T$-dependence of $m^*$.

From Eq.~(\ref{eq:se3}), a topological transition or crossover occurs at $U=U_c(T)$, where
\begin{equation}
\label{eq:uc}
U_c(T)=\sqrt{-m/f_z(T)}.
\end{equation}
Whether phonons favor topological or non-topological phases depends on the band parameters of the perfect crystal. 
If $t^2>\gamma^2$, ${\rm sgn}(f_z)=-{\rm sgn}(t)$ and phonons tend to drive the system into the topological insulator phase.
If $t^2<\gamma^2$, phonons instead stabilize the non-topological insulating phase~\cite{cave}.
Since $|f_z(T)|$ is a monotonically decreasing function of $x$, with $f_z\to 0$ as $x\to\infty$, phonons are more likely to alter the band topology in materials with slower Dirac fermions (smaller $\lambda$).

Some reasonable parameter values are~\cite{zhang2009,huang2008} $t\simeq 0.25\,{\rm eV}$, 
$\gamma\simeq 0.15\,{\rm eV}$, $\lambda\simeq 0.18 {\rm eV}$, $C\simeq 35\,{\rm eV}$, 
$\rho\simeq 7800\,{\rm kg/m}^3$, $c_s\simeq 1.7\, {\rm km/s}$ and $a\simeq 1\,{\rm nm}$. 
Then, $U\simeq 70\,{\rm meV}$ and  $U_c(0)\simeq 34\sqrt{m [{\rm meV}]}$.  
Accordingly, phonons may drive a topological quantum phase transition in narrow-gap semiconductors ($E_g\lesssim 10\, {\rm meV}$).
Moreover, because $U_c(T\gg\omega_c)\ll U_c(0)$, a material that is topologically trivial at $T=0$ may develop topological surface states at higher temperature.

\begin{figure}
\begin{center}
\includegraphics[scale=0.3]{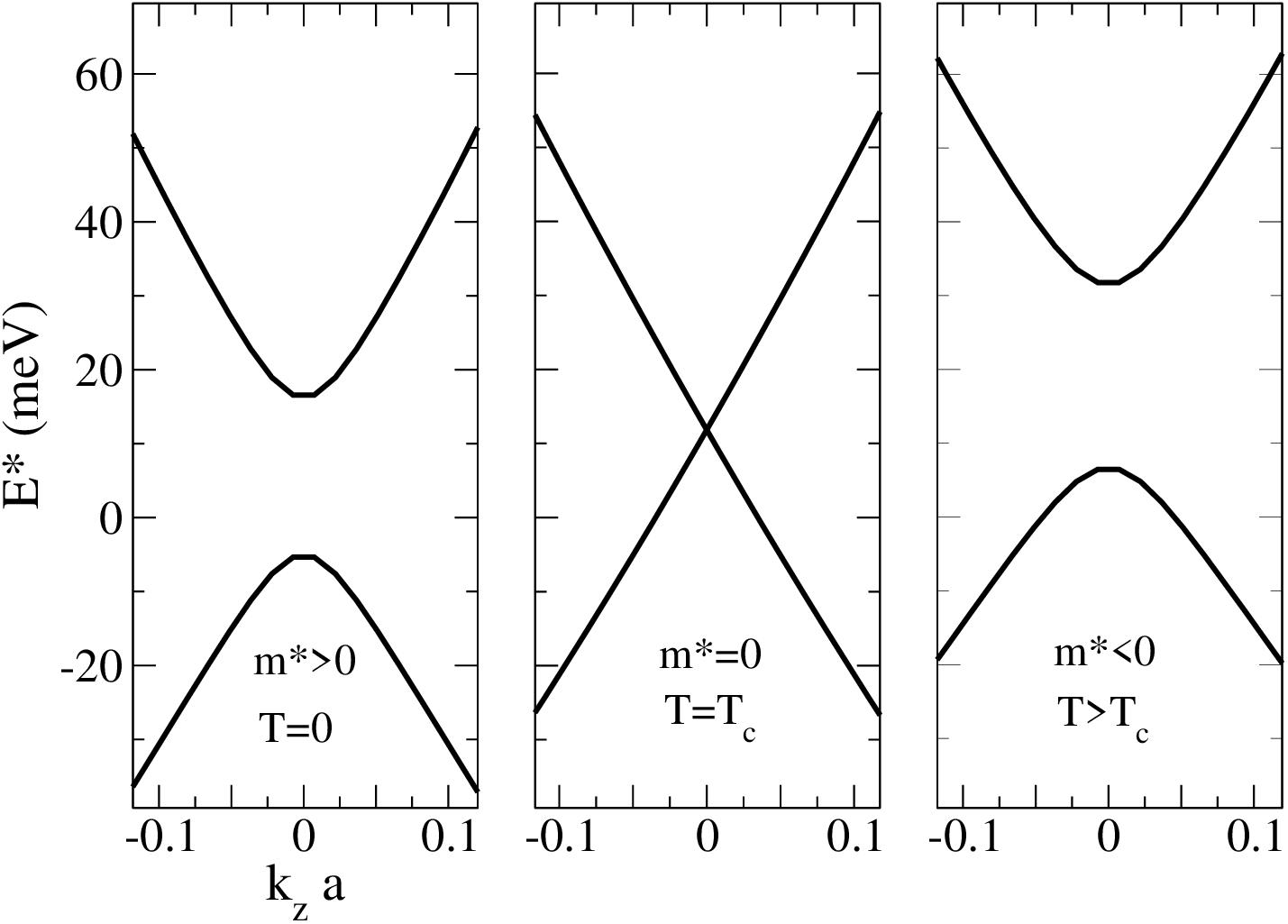}
\caption{Dressed bulk quasiparticle dispersion along [001], obtained from Eqs.~(\ref{eq:se}),~(\ref{eq:heff}) and (\ref{eq:model}). 
Parameter values are $U\simeq 57\, {\rm meV}$, $\gamma=0.15\, {\rm eV}$, $\lambda= 0.18\, {\rm eV}$, $t= 0.25\, {\rm eV}$, $\omega_c=8\, {\rm meV}$, $\omega_{\bf q}=\omega_c/(1.24\pi) a\,q$ and $m=20\, {\rm meV}$ (where $1.24 \pi/a$ is the radius of the Debye sphere). 
Because electron-phonon interactions preserve lattice and time-reversal symmetries, dressed bands are two-fold degenerate.
Left: $T=0$ (topologically trivial insulator). Middle: $T\simeq 0.7\,\omega_c$ (topological crossover temperature). Right: $T=1.2\,\omega_c$ (inverted regime). The band broadening $|\Gamma_{{\bf k}\alpha}|$ (not shown) grows with $T$, but remains $\lesssim 1\,{\rm meV}$ throughout the three panels.}
\label{fig:inversion}
\end{center}
\end{figure}

Aside from the Dirac mass, electron-phonon interactions renormalize the Fermi energy, $\epsilon_F^*(T)\simeq\epsilon_F-\Sigma_0(0)$~\cite{other}.
Though not explicitly, $U_c$ in Eq.~(\ref{eq:uc}) depends on $\epsilon_F^*$ implicitly because screening (ignored above) leads to $g_{\bf q}\to g_{\bf q}/\kappa({\bf q})$, where $\kappa({\bf q})\simeq 1+q_{TF}^2/q^2$ and $q_{TF}$ is the Thomas-Fermi wavevector. 
Therefore, the observation of phonon effects on topological properties may be contingent on having low bulk doping ($q_{TF}\ll k_c$).
In view of this, electrically gated thin films are the preferred venue to tune the chemical potential inside the bulk gap and to observe transport and thermodynamic signatures of phonon-induced surface states~\cite{pho}.

The thermal topological crossover introduced above is meaningful only if the quasiparticle band broadening in the vicinity of TRIM remains small compared to $|m^*|$.
The reduced phase space for scattering near $k=0$ helps keep the broadening small.
For example, it follows from inspection that there exists a finite window, 
$1\lesssim U^2 |f_z(T)|/m\lesssim 2/(1+\gamma/t)$,
 for which quasiparticle bands are inverted and yet sharp ($\Gamma_{{\bf 0}\alpha}=0$) at the same time. 
Here we have neglected the broadening due to e.g. impurities, bulk-surface scattering and electron-electron scattering.
The latter two are suppressed at $k\simeq 0$ insofar as $|m^*|> (\omega_c,T)$.

Along similar lines, the topological surface states derived from $G$ are hardly detectable by ARPES unless $|E_g^*|\simeq 2|m^*|\gg T$.
Whether this condition is met or not depends on material parameters, because $|m^*|\propto T$ at high temperature (c.f. Eq.~(\ref{eq:fz1})).
In our toy model, using the parameter values enlisted above, $|E_g^*|\gg T$ translates into $\omega_c\ll 25\,{\rm meV}$, which can only be satisfied in soft Dirac materials with low Debye temperatures.
This upper bound on $\omega_c$ becomes larger (and thus easier to fulfill) when $\lambda$ is smaller and/or when $\gamma$ is closer to $t$.

\begin{figure}
\begin{center}
\includegraphics[scale=0.3]{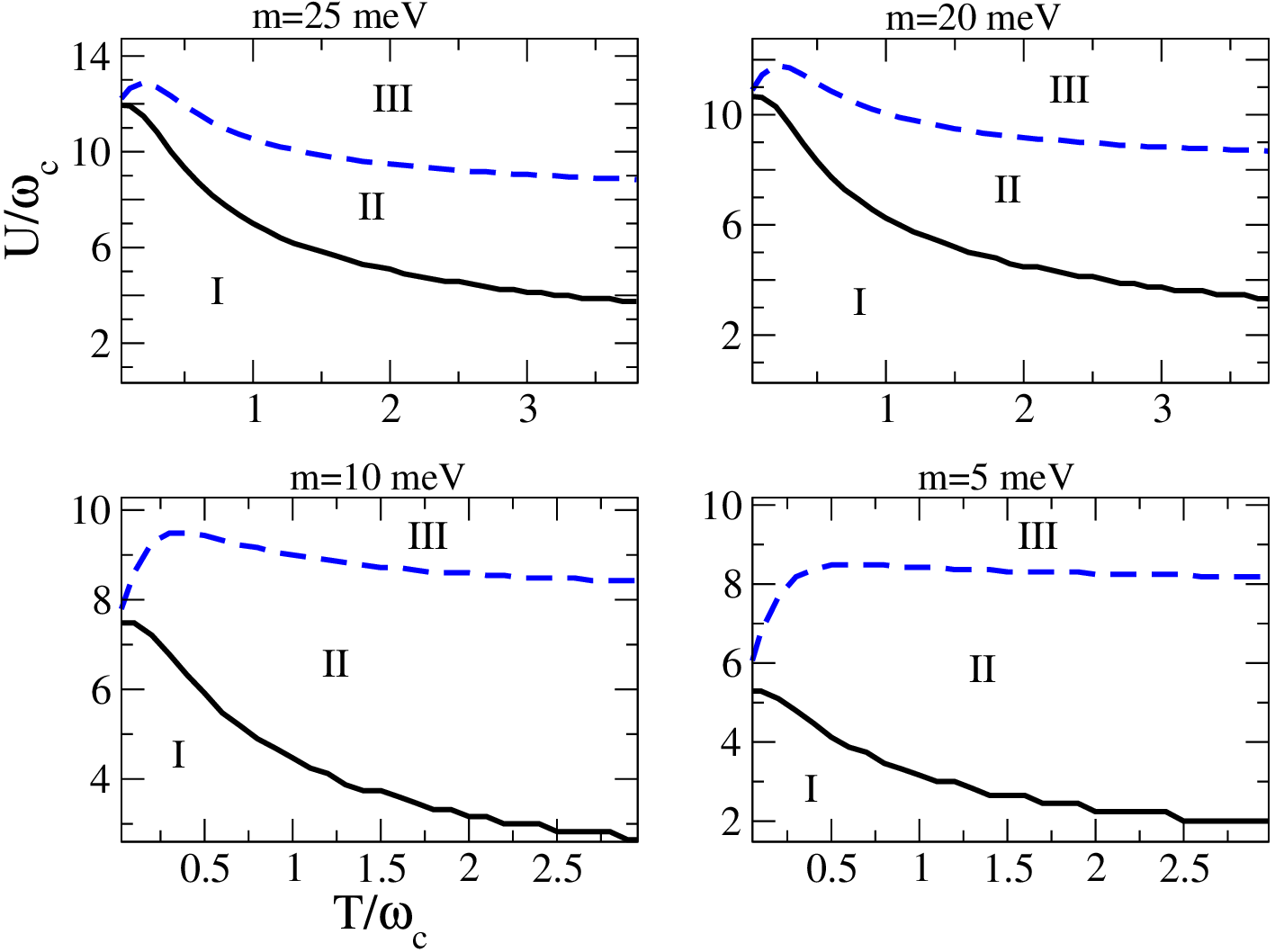}
\caption{Phonon-induced topological quantum phase transition (at $T=0$) and topological crossover (at $T>0$) for different values of the bare energy gap.
The parameters $(U,\gamma,\lambda,t,\omega_c)$ are the same as in Fig.~\ref{fig:inversion}.
I: non-topological insulator ($m^*>0$); II: ``crossover'' regime ($m^*<0$ and $|m^*|<4 T$); III: topological ``insulator'' ($m^*<0$ and $|m^*|>4 T$).
Because $\omega_c$ is fixed, the dashed lines in all the graphs converge to the same high-$T$ asymptote (c.f. Eq.~(\ref{eq:fz1})).}
\label{fig:crossover}
\end{center}
\end{figure}

Next, we proceed to solve the full lattice model numerically.
On one hand, Fig.~\ref{fig:inversion} displays the dressed quasiparticle spectrum, illustrating a band inversion at $T>0$. 
On the other hand, Fig.~\ref{fig:crossover} shows a topological crossover as a function of temperature.
The topological ``insulator'' at $T>0$ is defined as a phase where $m^*<0$ and $|E_g^*|\gtrsim 8 T$, the latter attribute being rather arbitrary.
For many semiconductors, $\partial E_g^*/\partial T\sim 1-10$ at high $T$~\cite{huang2008,allen1981,giustino2010}; hence it is conceivable that {\em some} real Dirac insulators exist such that $|E_g^*|\gtrsim 8 T$ deep in the inverted regime.
In the ``crossover'' region, where surface states are emerging, $m^*<0$ but $|E_g^*|\lesssim 8 T$.
In order to have a trivial insulator at $T=0$ turn into a topological ``insulator'' at high $T$, it helps if $m/\omega_c$ is larger because (i) at $T=0$, $U_c$ increases for larger $m$ (irrespective of $\omega_c$) and (ii) at high $T$, $|m^*|$ grows faster with $T$ for smaller $\omega_c$ (irrespective of $m$).

{\em Phonon-induced topological transitions in semimetals.--}
A simple lattice model for a 3D Dirac semimetal is
\begin{equation}
\label{eq:ws}
h({\bf k})=h^{(0)}({\bf k})+\Delta\sigma^z {\bf 1}_2,
\end{equation}
where $\Delta>0$ is an exchange field due to magnetic order that may originate e.g. from bulk dopants in a Dirac insulator.
When $m\in(-\Delta,\Delta)$, the energy spectrum of $h$ contains a pair of Weyl nodes of opposite chirality separated in momentum space (along [001]): the system is a Weyl semimetal (WS) and has Fermi arcs~\cite{ws} at the sample boundaries.
For $|m|\geq\Delta$, the two Weyl nodes merge and the crystal turns into an insulator.
Here we show that lattice vibrations, which preserve translational invariance only on average, can destroy or create Weyl nodes both at $T=0$ and $T>0$. 

\begin{figure}
\begin{center}
\includegraphics[scale=0.3]{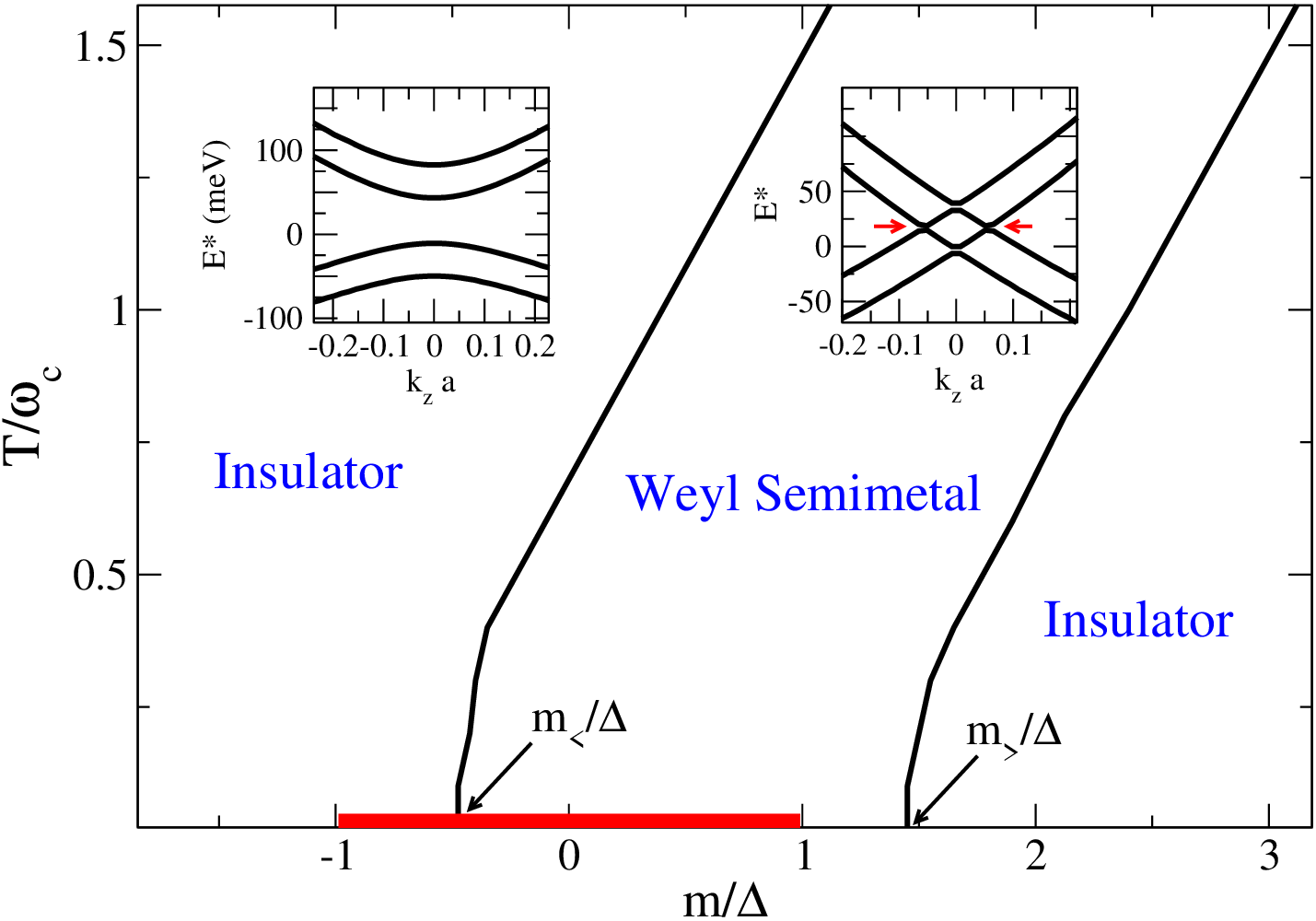}
\caption{Phase diagram for Eq.~(\ref{eq:ws}) in presence of electron-phonon interactions. 
The parameters $(U,t,\gamma,\lambda,\omega_c)$ take the same values as in Fig.~\ref{fig:inversion}, with $\Delta=20\, {\rm meV}$.
For simplicity we have neglected any $T$-dependence of $\Delta$ by assuming that the magnetic order producing it is robust up to $T\sim\omega_c$.
When $m>m_>$, increasing $T$ may induce a WS phase; for $m\in(m_<,m_>)$, increasing $T$ destroys the WS phase.
When $m<m_<$, the system is insulating for all $T$.
The thick horizontal line segment is the window for the WS phase in absence of electron-phonon interactions and at $T=0$. 
Insets: representative quasiparticle energy spectra for the insulating and WS phases, along [001]. 
Arrows indicate nodes with nonzero topological invariant.} 
\label{fig:sm}
\end{center}
\end{figure}

We evaluate $E^*_{{\bf k}\alpha}$ by combining Eqs.~(\ref{eq:se}),~(\ref{eq:heff}) and (\ref{eq:ws}).
The resulting phase diagram (Fig.~\ref{fig:sm}) evidences that, in presence of phonons, (i) $\Delta$ remains nearly unchanged, (ii) the WS phase is stable for $m^*(T)\in(-\Delta,\Delta)$, and (iii) the Weyl nodes occur away from zero energy due to $\epsilon_F\to\epsilon^*_F$.
These generic observations are likely to hold for more elaborate models than Eq.~(\ref{eq:ws}).
In addition, Fig.~\ref{fig:sm} proves that a material which is insulating at $T=0$ may turn into WS at nonzero temperature.
In order to observe this thermally-induced WS phase, the temperature at which the magnetic order giving rise to $\Delta$ dissolves must not be low compared to the Debye temperature.

{\em Conclusions--}
We have given a proof-of-principle for the impact of electron-phonon interactions on the band topology of insulators and semimetals, with an emphasis on thermally driven topological crossovers.
Other bosonic baths may likewise play a role in inducing or destroying topological phases in thermal equilibrium.

A necessary future task will consist of searching for real Dirac materials with appropriate band parameters to realize the qualitative effects discussed herein.
Generally, the more promising candidates will be soft crystals with strong electron-phonon coupling and a small zero-temperature bandgap that is nonetheless larger than the Debye temperature.
Specifically, BiTl(S$_{1-\delta}$ Se$_\delta$)$_2$ seems interesting because both the magnitude and sign of its bandgap are tunable~\cite{xu2011}. 
However, it is not yet known how strong the phonon-induced bandgap renormalization is in this material.
Should BiTl(S$_{1-\delta}$ Se$_\delta$)$_2$ become ferromagnetic upon magnetic doping, there would also be an opportunity to alternate between insulating and Weyl semimetallic phases by tuning $\delta$ and the temperature. 


I acknowledge valuable interactions with  V. Albert, M. Franz, L. Glazman, J. Moore, J.D. Sau and J. Vayrynen. 
This project has been funded by Yale University, and has also benefited from the kind hospitality of the Aspen Center for Physics through NSF Grant No. 1066293.


\begin{thebibliography}{30}
\bibitem{expansion} We keep the volume of the crystal fixed and thus neglect the variations in the electronic structure that result from the thermal expansion of the lattice. 
These variations are typically small compared to the electron-phonon contribution calculated here.
\bibitem{freq} This is a good approximation insofar as $|\omega|$ is small compared to the electronic bandwidth.
\bibitem{ee} Using Eq.~(\ref{eq:model}) and the Hartree-Fock approximation, the renormalized Dirac mass due to electron-electron interactions is $m^*\simeq m -1/(4 \pi^2) \int_{0}^{\pi/a} dq (M_q/E_q) \, e^2/(4\pi\epsilon) \tanh(E_q/2 T)$, where $e$ is the electron's charge and $\epsilon$ is the dielectric constant.
The $T$-dependent part of this expression is small relative to its $T$-independent part when the temperature is low compared to the typical ($\sim {\rm O}(1 {\rm eV})$) electronic bandwidth.
Unlike with phonons, the influence of purely electronic interactions on the Dirac mass decreases as temperature increases.
\bibitem{cave} Note, however, that for $t^2<\gamma^2$, Eq.~(\ref{eq:model})  models either an indirect-gap semiconductor or a metal.
\bibitem{other} We find the phonon-induced renormalizations of $t,\lambda$ and $\gamma$ to be very weak.
This fact, along with the approximate frequency- and $m$-independence of the self-energy, implies that a self-consistent Born approximation for the electron-phonon coupling will not qualitatively change the results of this work.
\bibitem{pho} Such films are typically thinner than the mean free path of phonons; hence, their phonon spectra will show deviations from the bulk spectra assumed in this work. These deviations will result in quantitative changes for the topological crossover temperature.
\end{thebibliography}
\end{document}